# Ultrahigh electrostrain > 1 % in lead-free piezoceramics: A critical review

Gobinda Das Adhikary, Digivijay Narayan Singh, Getaw Abebe Tina, Gudeta Jafo Muleta, Rajeev Ranjan

Department of Materials Engineering, Indian Institute of Science, Bengaluru-560012, India.

## Abstract

Recently, a series of reports showing ultra-high electrostrain (> 1 %) have appeared in several Pb-free piezoceramics. The ultrahigh electrostrain has been attributed exclusively to the defect dipoles created in these systems. We examine these claims based on another report https://arxiv.org/abs/2208.07134 which demonstrated that the measured electric field driven strain increased dramatically simply by reducing the thickness of the ceramic discs. We prepared some representative Pb-free compositions reported to exhibit ultrahigh strain and performed electrostrain measurements. We found that these compositions do not show ultrahigh electrostrain if the thickness of the discs is above 0.3 mm (the disc diameters were in the range 10- 12 mm diameter). The ultrahigh strain values were obtained when the thickness was below 0.3 mm. We compare the electrostrain obtained from specimens designed to exhibit defect dipoles with specimens that were not designed to have defect dipoles in $Na_{0.5}Bi_{0.5}TiO_3$ (NBT) and $K_{0.5}Na_{0.5}NbO_3$ (KNN) -based lead-free systems and could obtain much higher strain levels (4- 5 %) in the defect dipole free piezoceramics in the small thickness regime. Our results do not favor the defect dipole theory as the exclusive factor for causing ultrahigh strain in piezoceramics. A new approach is called for to understand the phenomenon of ultrahigh electrostrain caused by the thickness reduction of piezoceramic discs.

*rajeev@iisc.ac.in

There is great interest in piezoceramics exhibiting large electrostrain. Earlier, large electrostrain > 1 % were thought to be possible only in suitably oriented single crystals of morphotropic phase boundary compositions of ferroelectric systems [1]. Polycrystalline piezoceramics were considered to exhibit considerably smaller electrostrain (in the range 0.2 – 0.4 %) [2]. In 2015, Liu and Tan [3] reported an electrostrain of 0.7 % in a $Na_{0.5}Bi_{0.5}TiO_3$-derived polycrystalline piezocermic disc. Narayan et al. reported electrostrain of ~ 1 % in a polycrystalline piezoceramic of a pseudoternay $BiFeO_3$-$PbTiO_3$-$LaFeO_3$ [4]. Recently, Adhikary and Ranjan [5] reported that when the thickness of piezoceramics discs (10-12 mm in diameter) is reduced below 0.5 mm, the measured electrostrain values shoot up sharply. Electrostrain values of ~ 2 % were observed even in normal piezoceramics like PZT and modified $BaTiO_3$ [5]. Soon after the publication of this report, there is a spurt in publications showing large electrostrain > 1 % in lead-free piezoceramics [6 - 12]. A common theme in all these reports is that the exceptionally large electrostrain is caused primarily by the defect dipoles created in the system. Different groups have envisaged different chemical modification strategies to create defect dipoles [6-12]. Feng et al. [9] reported that when oxygen vacancies are created by controlled volatilization of the atomic species in $Na_{0.5}Bi_{05}TiO_3$, the electrostrain can be increased to ~ 0.8 % at room temperature. When heated close to the depolarization temperature ~ 220 ºC, the strain reaches 2.3 % [9]. They introduced a term "hetrostrain" to describe the observation of negative strain on the negative field and positive strain on the positive cycle, during field cycling of the piezoceramic. Given that this is merely a defective (oxygen deficient) NBT, the strain of 0.8 % is considerably large. Before this report, the maximum electrostrain of 0.7 % was reported at room temperature in a complex NBT-based system, namely $(Bi_{1/2}(Na_{0.84}K_{0.16})_{1/2})_{0.96}Sr_{0.04})(Ti_{0.975}Nb_{0.025})O_3$ [3]. Luo et al. [10] introduced oxygen vacancies via modifying NBT with a hypothetical perovskite $BaAlO_{2.5}$ and reported an electrostrain of 1.12 % (at 100 kV/cm) at room temperature. Still higher strain 1.68 % was reported Jia et al.[11] in $(Bi_{0.5}Na_{0.5})_{0.94}Ba_{0.06}Ti_{1-x}(Zr_{0.50}Sb_{0.4}\square_{0.1})_xO_3$ and $(Bi_{0.5}Na_{0.5})_{0.94}Ba_{0.06}Ti_{1-x}(Sn_{0.50}Sb_{0.4}\square_{0.1})_xO_3$ (x=0.01 and 0.02) polycrystalline piezoceramics.

That defect dipole creation is the critical factor for achieving such large electrostrain in piezoceramics has also been argued in other Pb-free piezoelectric systems. Li et al. [12] reported electrostrain ~ 1 % at room temperature in a complex derivative of $BiFeO_3$, i.e., $0.695BiFeO_3$-$0.3BaTiO_3$-$0.005Bi(Zn_{0.5}Ti_{0.5})O_3$. The authors attributed the extraordinary electrostrain strain to polar nano regions and the coexistence of phases.

Among the KNN-based piezoceramic derivatives, Huangfu *et al*., [6] reported electrostrain of 1.05 % at room temperature in Sr-modified KNN and attributed it to the presence of $V'_{K/Na}$ - $V_O$ defect dipoles. In subsequent work, the electrostrain was reported to increase to 1.35 % in textured ceramic of the same composition [7].

Given the observation of Adhikary and Ranjan [5], it is essential to re-look if the ultrahigh electrostrain reported in some of the lead-free compositions owes their origin to the defect dipoles introduced deliberately by the design of compositions, or if it is a consequence of reduced disc thickness. The significant interest in high electrostrain materials makes this issue crucial to resolve. Here, we have examined this issue critically. We chose one system based on KNN and another system based on NBT. We synthesized a Sr-modified KNN lead-free piezoceramics for which electrostrain ~ 1.05 % has been reported [6]. For the NBT-based lead-free piezoelectrics, we followed Luo et al. [10], who reported that oxygen vacancies introduced by modifying NBT with hypothetical perovskite $BaAlO_{2.5}$ cause the electrostrain to increase to 1.12 %. The specimens were prepared using the conventional solid state sintering method. Electrostrain and polarization measurements were carried out on ~ 95 % dense sintered discs. For a given composition, measurements were performed on circular discs (diameter 10-12 mm) of different thicknesses in the thickness regime 1 mm – 0.2 mm. The strategy of thickness reduction was followed as reported in ref [5]. Polarization and electrostrain measurements were performed with the modular set up of Radiant ferroelectric measurement system at 1 Hz on discs electroded with silver paint.

Following Huangfu *et al*.[6], we synthesized $K_{0.5(1-x)}Na_{0.5(1-x)}Sr_x NbO_3$ with x=0.02 (KNNS2). The bipolar polarization, electrostrain as well as the unipolar electrostrain of this composition is shown in Fig. 1.

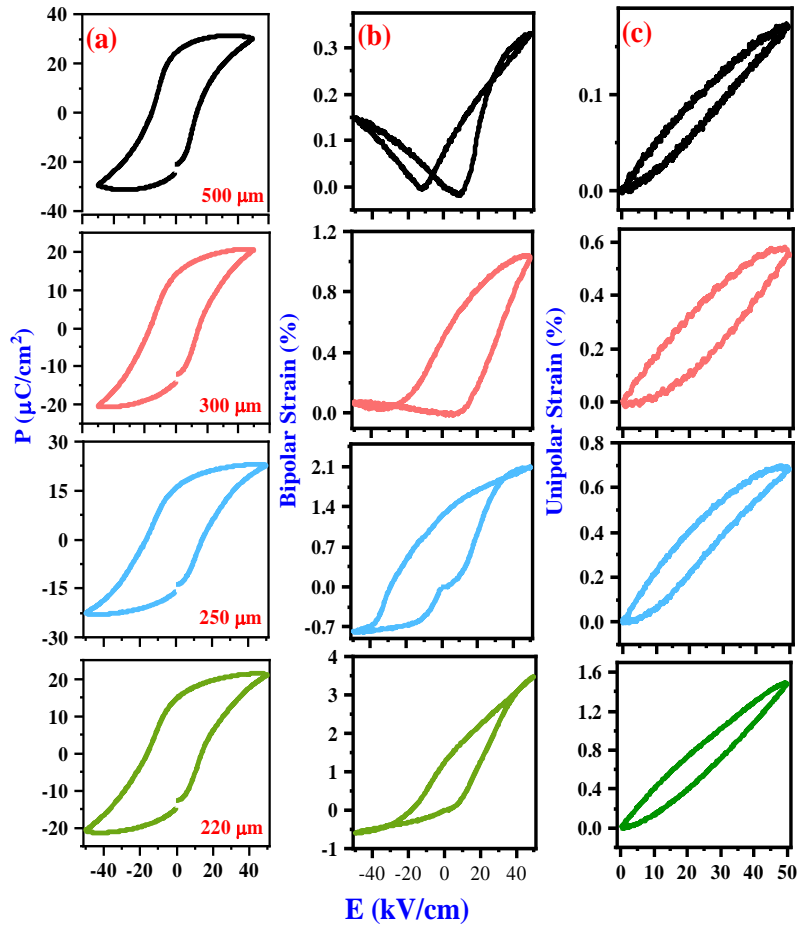

**Fig. 1** Ferroelectric and electrostrain measurements of KNNS2. (A) Electric polarization as a function of the electric field, (b) Strain-electric (S-E) field (Bi-polar) curves, and (c) Strain-electric (S-E) field (Mono-polar) curves for 500 μm, 300 μm, 250 μm, and 220 μm thick pellets, recorded at room temperature.

The maximum bipolar electrostrain measured on the 0.5mm thick disc is ~0.35 % at a field of + 50 kV/cm. The corresponding unipolar electrostrain is nearly half this value (0.18 %). As the thickness is reduced, two things happen: (i) the bipolar strain loop becomes increasingly asymmetrical. The strain on the negative cycle decreases, and that on the positive side increases. For thickness below 0.3mm, the strain is negative in the negative cycle and positive in the positive cycle. Such shapes were called hetrostrain by Feng *et al*. [9] and were attributed to oxygen vacancies. The strain in the positive cycle reaches ~3.5 % when measured on 0.22 mm disc. Here we see that the shape change from "butterfly loop" to "hetrostrain loop" is merely a consequence of thickness reduction. The unipolar electrostrain (which is the most important parameter for any actuator application), is merely 0.18 % when measured on 0.5 mm thick disc. It could reach 1.4 % kV/cm when the thickness was reduced to 0.22 mm. This value exceeds the one reported by Huangfu *et al*. [6].

Having proven above that the measured ultrahighstrain on KNNS2 is not a compositional (defect dipole) effect but a direct consequence of the reduction in the disc thickness, we also investigated this effect in conventional MPB compositions of KNN-derivatives. KNN exhibits orthorhombic (Amm2) ferroelectric phase at room temperature. Li modification induces the tetragonal (P4mm) phase [13]. We synthesized a composition $K_{0.5(1-x)}Na_{0.5(1-x)}Li_xNbO_3$ with x=0.06 (KNNL6) corresponding to a P4mm-Amm2 boundary and measured electrostrain as a function of thickness for this composition, Fig. 2.

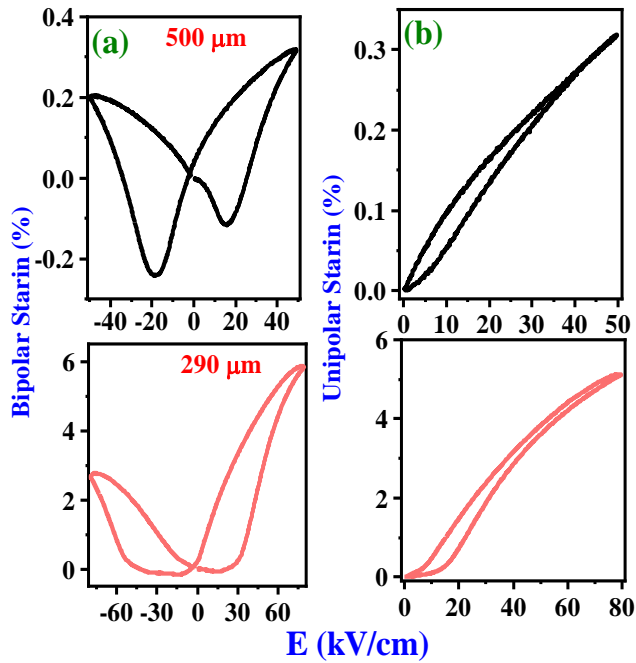

**Fig. 2** (a) Bipolar and (b) unipolar electrostrain measured on 0.5 mm and 0.29 mm thick pellets of KNNL6 system.

Similar to KNNS2, the strain develops asymmetry in bipolar measurements on thickness reduction. The maximum strain at 50 kV/cm measured on the 0.5 mm thick disc is ~0.3 %, which is almost similar to the value obtained on KNNS2 of the same thickness. In fact, the unipolar electrostrain ~ 0.3 % of this MPB composition is better than the unipolar electrostrain of the KNNS2 composition. With the reduction in the thickness of the disc, the dielectric breakdown strength of the ceramic improved we could apply higher fields (80 kV/cm). As evident from Fig. 2, the maximum electrostrain in bipolar and unipolar measurements on 0.29 mm disc of KNNL6 is ~ 6 % and ~ 5 % (at 80 kV/cm), respectively. Given that, unlike KNNS2, the MPB composition of Li-modified KNN was not designed to possess point defects, the extraordinary electrostrain of KNNL6 suggests that compositional design approaches aimed at the creation of defect dipoles are not important for obtaining ultrahigh electrostrain in piezoceramics.

We also synthesized oxygen-deficient NBT-based lead-free piezoceramics to examine the hypothesis of oxygen deficiency playing crucial role in the ultrahigh measured strain in NBT-based systems. We followed Luo et al. [10], who reported that oxygen-deficient (1-x)NBT-(x)BaAlO$_{2.5}$ piezoceramics show electrostrain of 1.12 % at room temperature. Here BaAlO$_{2.5}$ is treated as a hypothetical perovskite which, when alloyed with a real stoichiometric perovskite, will yield oxygen-deficient perovskite. We synthesized the same composition x=0.06 of this system as reported in ref [10] and measured the bipolar and unipolar electrostrain for different thicknesses. For this composition, the thick and thin discs could easily survive an electric field upto 80 kV/cm. As shown in Fig. 3, the 0.7 mm thick disc shows bipolar and unipolar electrostrain of 0.5 % and 0.35 %, respectively, at 80 kV/cm. These values are significantly less than what was reported by Luo *et al.*(~1.12 % at 100 kV/cm) [10].

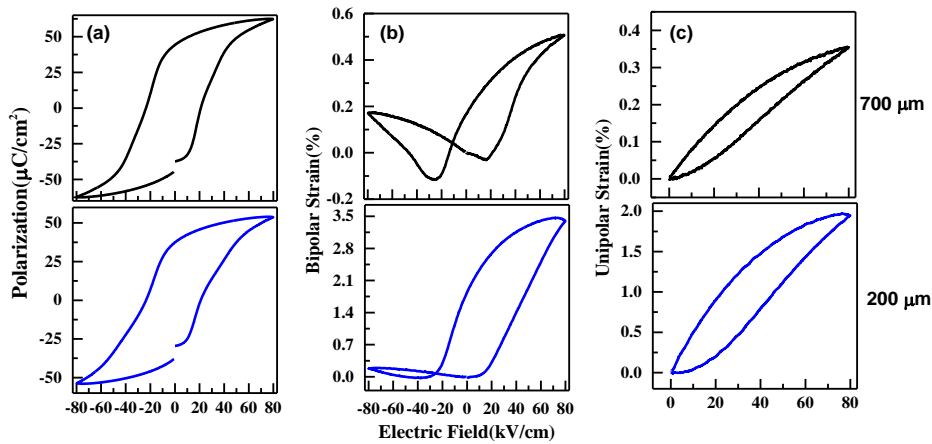

**Fig. 3:** (a) P-E loops, (b) bipolar strain and (c) unipolar strain loops measured on 700μm and 200μm thick pellets of Na$_{0.5}$Bi$_{0.5}$TiO$_3$-0.06BaAlO$_{2.5}$ .

However, when the thickness of the disc was reduced to 0.2mm, the measured bipolar strain increased enormously to 3.5 %, and the unipolar strain increased to ~ 2 %. These values are much higher than what was reported by Luo *et al*. [10]. These observations confirm that the disc's reduced thickness is primarily responsible for the ultrahigh measured electrostrain in Na$_{0.5}$Bi$_{0.5}$TiO$_3$-0.06BaAlO$_{2.5}$.

For direct comparison, we also investigated NBT derivatives without deliberately created defect dipoles. Here we present results on two systems, namely 0.62Na$_{0.5}$Bi$_{0.5}$TiO$_3$-0.28SrTiO$_3$ (Fig. 4) and 0.62Na$_{0.5}$Bi$_{0.5}$TiO$_3$-0.20K$_{0.5}$Bi$_{0.5}$TiO$_3$-0.08NaNbO$_3$ (Fig. 5). Both the compositions correspond to the ergodic-non ergodic relaxor state boundary at room temperature. The 0.7 mm thick 0.62Na$_{0.5}$Bi$_{0.5}$TiO$_3$-0.28SrTiO$_3$ disc shows bipolar and unipolar electrostrain of 0.25 % at 50 kV/cm. However, when the thickness is reduced to 0.2mm, the bipolar strain in the positive cycle reaches 5 %. The unipolar strain reaches 4.5 %

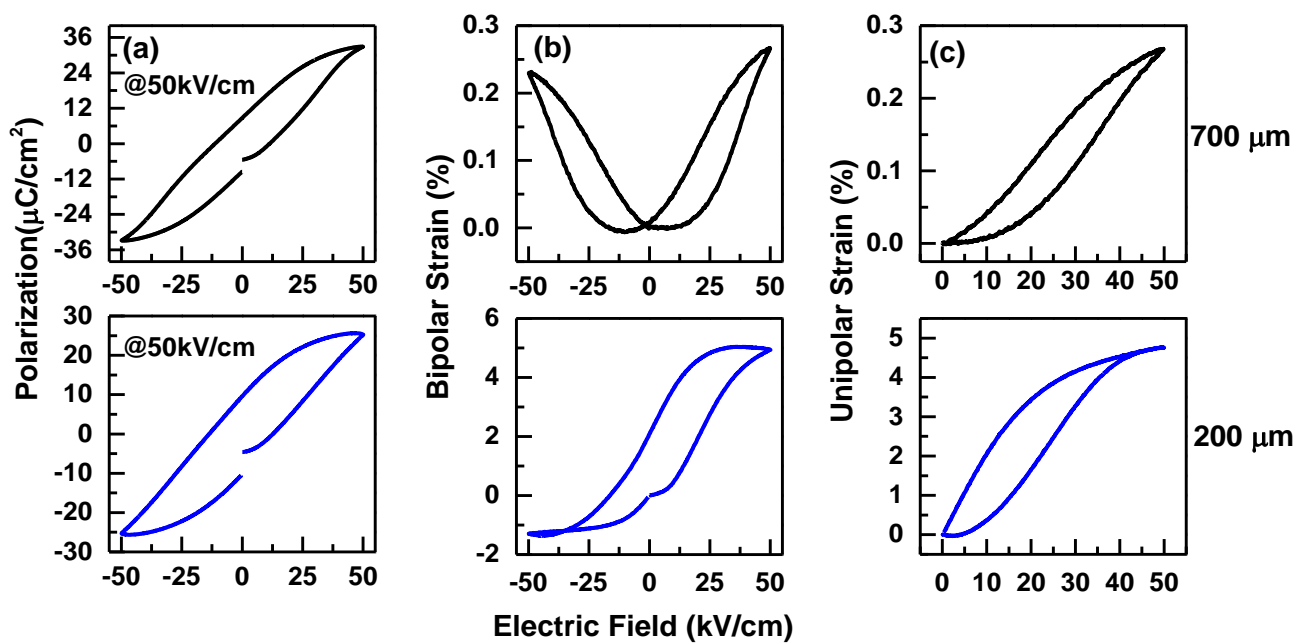

**Fig. 4:** (a) P-E loops, (b) bipolar strain and (c) unipolar strain loops measured on 700μm and 200μm thick pellets of $0.62Na_{0.5}Bi_{0.5}TiO_3$-$0.28SrTiO_3$ system.

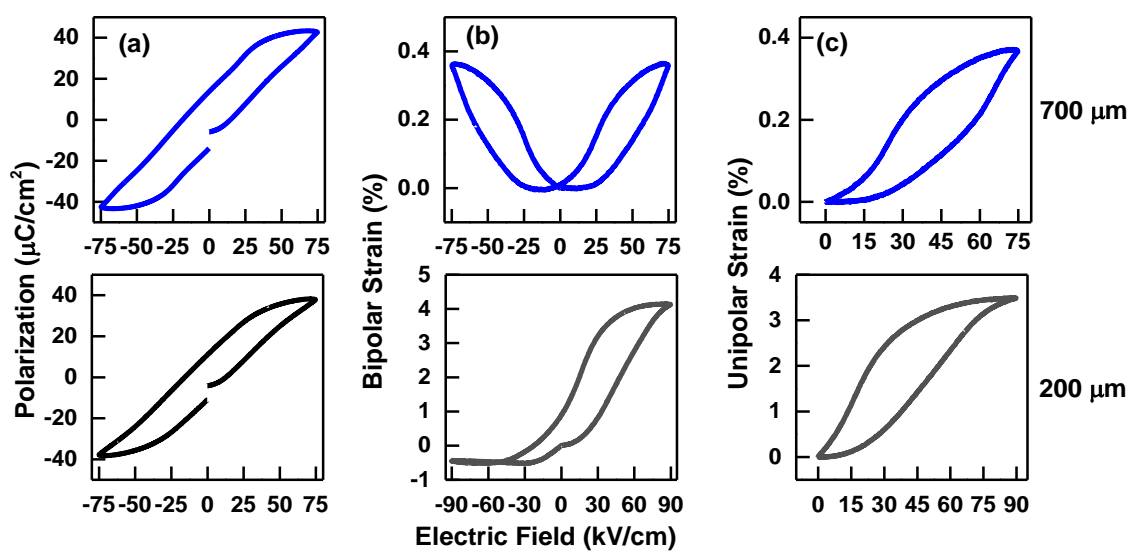

**Fig. 5 :** (a) P-E loops, (b) bipolar strain and (c) unipolar strain loops measured on 700μm and 200μm thick pellets of $0.68Na_{0.5}Bi_{0.5}TiO_3$-$0.20K_{0.5}Bi_{0.5}TiO_3$-$0.08NaNbO_3$ system.

For the other system $0.68Na_{0.5}Bi_{0.5}TiO_3$-$0.20K_{0.5}Bi_{0.5}TiO_3$-$0.08NaNbO_3$, the 0.7 mm thick disc exhibits strain of 0.4 % at 80 kV/cm. It shoots to 4 % and 3.5 % in bipolar and unipolar measurements, respectively. Interesting to note that these values are significantly larger than the one obtained for the defect dipole containing system 0.94 NBT-0.06BaAlO$_3$ (ref [10], and Fig. 3).

Our results confirm that the ultrahigh strain (> 1 %) reported in some of the lead-free piezoceramics is not primarily due to the defect dipoles, as generally proposed. The measured ultrahigh electrostrain is a phenomenon rather associated with the reduction in the thickness of the piezoceramic discs. We show that much larger electrostrain can be obtained in systems that are not designed to have defect dipoles (as in the MPB composition of the Li-modified KNN system or NBT systems at the ergodic – nonergodic relaxor boundary). Given the huge effect of reduced thickness on the measured strain, the emphasis should be on understanding what thickness reduction does to make the piezoceramic discs exhibit such a large increase in the measured electrostrain. Care must be taken about possible artifacts in the measurements. Recently, a nominal strain of 0.7 % at 150 kV/cm was reported in non-perovskite piezoceramic ceramic $Bi_2WO_6$, and the authors attributed it to bending of the ceramic disc [14]. Such possibility need not be excluded in the perovskite-based piezoceramics.

Rajeev Ranjan acknowledges Science and Engineering Research Board (SERB), India and Indian Institute of Science (IISc) for supporting research on piezoelectric materials.

**References**

[1] S. E. Park, & T. R. Shrout, Ultrahigh strain and piezoelectric behavior in relaxor based ferroelectric single crystals. *J. Appl. Phys.* **82**, 1804-1811 (1997).

[2] J. Hao, W. Li, J. Zhai, & H. Chen H, Progress in high-strain perovskite piezoelectric ceramics, *Mat. Sci. & Engg. R Reports* **135**, 1-57 (2019).

[3] [9] X. Liu and X. Tan, *Giant Strains in Non-Textured (Bi 1/2 Na 1/2 )TiO 3 -Based Lead-Free Ceramics, Adv. Mater.* **28**, 574-578 (2016).

[4] B. Narayan, J. S. Malhotra, R. Pandey, K. Yaddanapudi, P. Nukala, B. Dkhil, A. Senyshyn, R. Ranjan, Electrostrain in excess of 1 % in polycrystalline piezoceramics. *Nat. Mater*. 17, 427-431 (2018).

[5] Gobinda Das Adhikary and Rajeev Ranjan, *Ultrahigh measured unipolar strain > 2 % in polycrystalline bulk piezoceramics: Effects of disc dimension,* arxiv.org/abs/2208.07134

[6] Geng Huang fu, Kun Zeng, Binquan Wang, Jie Wang, Zhengqian Fu, Fangfang Xu, Shujun Zhang, Haosu Luo, Dwight Viehland, Yiping Guo, *Giant electric field–induced strain in lead-free piezoceramics*, Science 378, 1125–1130 (2022)

[7] Lixiang Lai, Bin Li, Shuo Tian, Zhihao Zhao, Shujun Zhang, Yejing Dai, *Giant electrostrain in lead-free textured piezoceramics by defect dipole design*, Adv. Mater 2300519 (2023) https://doi.org/10.1002/adma.202300519

[8] Binquan Wang, Geng Huangfu, Zhipeng Zheng, and Yiping Guo, *Giant Electric Field-Induced Strain with High Temperature-Stability in Textured KNN-Based Piezoceramics for Actuator Applications*, Adv. Funct. Mater. 2023, 33, 221464; DOI: 10.1002/adfm.202214643

[9]Wei Feng, Bingcheng Luo, Shuaishuai Bian, Enke Tian, Zili Zhang, Ahmed Kursumovic, Judith L. MacManus-Driscoll, Xiaohui Wang & Longtu Li, *Heterostrain-enabled ultrahigh electrostrain in lead-free piezoelectric*, Nature Communications | ( 2022) 13:5086; https://doi.org/10.1038/s41467-022-32825-9

[10] Huajie Luo, Hui Liu, Houbing Huang, Yu Song, Matthew G. Tucker, Zheng Sun, Yonghao Yao, Baitao Gao, Yang Ren, Mingxue Tang, He Qi, Shiqing Deng, Shujun Zhang, Jun Chen, *Achieving giant electrostrain of above 1% in (Bi,Na)TiO3-based lead-free piezoelectrics via introducing oxygen-defect composition*, Sci. Adv. 9, 7078 (2023)

[11] Yuxin Jia, Huiqing Fan, Ao Zhang, Han Wang, Lin Lei, Qifeng Quan, Guangzhi Dong, Weijia Wang, Qiang Li, *Giant electro-induced strain in lead-free relaxor ferroelectrics via defect engineering*, Journal of the European Ceramic Society 43 (2023) 947–956

[12] W. Li, C. Zhou, J. Wang, C. Yuan, J. Xu, Q. Li , G. Chen, J. Zhao, G. Rao, Giant electro-strain nearly 1% in BiFeO3-based lead-free piezoelectric ceramics through coupling morphotropic phase boundary with defect engineering, Materials Today Chemistry 26 (2022) 101237).

[13] E. Hollenstein, M. Davis, D. Damjanovic, N. Setter, *Piezoelectric properties of Li- and Ta-modified (K$_{0.5}$Na$_{0.5}$)NbO$_3$ ceramics*, Appl. Phys. Lett. 87, 182905 (2005)

[14] Xiang He, Chen Chen, Lu Wang, Yunyun Gong, Rongmin Dun, Faqiang Zhang, Yanqiu Wu, Huarong Zeng, Yongxiang Li, Zhiguo Yi, *Giant electromechanical response in layered ferroelectrics enabled by asymmetric ferroelastic switching,* Materials Today 58, 48-56 (2022).